\documentclass[aps,pre,twocolumn,showpacs,%
preprintnumbers,amssymb]{revtex4-1}
\usepackage{graphicx}
\usepackage[colorlinks=true,linkcolor=blue,%
pagecolor=blue,citecolor=blue,%
urlcolor=blue,anchorcolor=black,%
bookmarksnumbered=true,%
bookmarksopen=true]{hyperref}
\usepackage{dcolumn}
\usepackage{bm}
\renewcommand{\frac}[2]{\displaystyle{#1 \over #2}}
\begin{document}
\title{New approach to measurement of the three dimensional crystallization front propagation velocity in strongly coupled complex plasma}
\author{D.~I.~Zhukhovitskii}
\affiliation{Joint Institute of High Temperatures, Russian Academy of Sciences, Izhorskaya 13, Bd.~2, 125412 Moscow, Russia}
\affiliation{Moscow Institute of Physics and Technology, Institutsky lane 9, Dolgoprudny, Moscow region, 141700 Russia}
\author{V.~N.~Naumkin} \email{naumkin@ihed.ras.ru}
\affiliation{Joint Institute of High Temperatures, Russian Academy of Sciences, Izhorskaya 13, Bd.~2, 125412 Moscow, Russia}
\author{V.~I.~Molotkov}
\affiliation{Joint Institute of High Temperatures, Russian Academy of Sciences, Izhorskaya 13, Bd.~2, 125412 Moscow, Russia}
\author{A.~M.~Lipaev}
\affiliation{Joint Institute of High Temperatures, Russian Academy of Sciences, Izhorskaya 13, Bd.~2, 125412 Moscow, Russia}
\author{H.~M.~Thomas}
\affiliation{Institut f\"{u}r Materialphsyik im Weltraum, Deutsches Zentrum f\"{u}r Luft  und Raumfahrt (DLR), M\"{u}nchener Str.\ 20, 82234 Wessling, Germany}
\date{\today}
\begin{abstract}
The PK-3 Plus laboratory onboard the International Space 
Station is used to form complex plasma with a liquidlike 
particle subsystem in metastable state and to observe the 
propagation of crystallization fronts corresponding to the 
surfaces of crystal domains. We propose the ``axis'' 
algorithm of solidlike particles identification, which 
makes it possible to isolate different domains and their 
surfaces as well. Determination of the three-dimensional front 
velocity is based on its definition implying that there exists a 
small area of the domain surface propagating along some line 
perpendicularly to it. The velocity measured in this way is an 
important characteristic of the plasma crystallization kinetics. 
It proves to be almost independent of time and the direction of front propagation and amounts to ca.\ 60--80\,$\mu {\mbox{m~s}}^{-1}$.
\end{abstract}
\maketitle

\section{\label{s1}INTRODUCTION}

A low-temperature plasma that contains dust particles 
typically in the range from tens of nanometer to thousands of 
micrometers is commonly called complex (or dusty) 
plasma \cite{1,2,3,5,6,9}. 
Dusty plasmas are present in astrophysical objects such as comets, protoplanetary discs, or planetary rings \cite{Goertz1989}.
Complex plasmas 
are specially prepared to study fundamental 
processes in the strong coupling regime on the most 
fundamental (kinetic) level, through the observation of 
individual microparticles and their interactions. In 
ground-based experiments, the microparticles are heavily 
affected by the force of gravity. Under microgravity 
conditions, e.g. on the International Space Station (ISS), 
gravity is compensated by the ISS orbital motion \cite{10,15,16,17,18,019,19,101}. 
Due to the high mobility of 
electrons, particles acquire a significant (macroscopic) 
negative electric charge. As a result, the particles are 
pushed out of the strong electric field region close to the 
electrodes. Thus, they can form expanded almost 
homogenous particle clouds in the bulk of the low-pressure 
gas discharge. Also, due to the large particle charge, such 
subsystem can form a three-dimensional (3D) complex plasma analog of a condensed state that has an ordered structure. We will term the particle subsystems ``liquidlike'' and ``solidlike'' (dust crystal) if the dust cloud is similar to a liquid or a solid, respectively. The dust crystal can undergo phase transitions similar to that in an equilibrium system, in particular, the melting--crystallization first-order transition \cite{1,101}.

Strongly coupled electron-ion plasmas are encountered in numerous physical and astrophysical situations, e.g., inertially confined laboratory plasmas, liquid metals, stellar and planetary interiors, and supernova explosions \cite{Chabrier1994}. 
Under certain conditions, one can expect the plasma crystallization process associated with the front propagation.

The most valuable information can be gained from 
observations of the phase transition kinetics. In 
ground-based experiment \cite{89}, growth of a crystal 
domain in the ``liquidlike'' dust cloud was observed. Based 
on the crystallization front visualization, the authors 
observed its propagation. However, an insufficient system 
homogeneity and the absence of 3D scanning did not make 
it possible to measure the ``true'' 3D front velocity. Instead, 
only one velocity component in the plane of the system 
light illumination was estimated on the order of magnitude. 
Note that the growth of solid phase nuclei in 
``supercooled'' colloidal systems \cite{100,102} is similar 
to the processes occurring in complex plasma.

We used the PK-3 Plus laboratory onboard the ISS 
\cite{18,92,69} to form a ``supercooled'' liquidlike 
microparticles system and to observe formation and growth 
of the crystal domains containing $\gtrsim 10^4$ particles 
each. Due to depth scans performed during the 
experiment, we can recover 3D coordinates of each 
microparticle. Then we use and compare different methods 
to isolate individual crystal domains and their surfaces.

The objective of this work is to measure the 3D 
crystallization front velocity. This task is complicated by 
two facts. First, the domain surface is quite irregular, and 
the intervals between successive depth scans are too long so 
that the displacement between corresponding points of the 
propagating front cannot be determined directly. Second, 
the finite scan velocity results in surface distortion, which 
means that the visible and true domain surfaces are 
different. We solve these problems by determination of a 
small flat area of the domain surface that propagates along 
some line and is normal to this line (direction of the 
rectilinear propagation). This enables one to measure the 
displacement of this area between successive scans and 
thus to determine the front velocity.

The paper is organized as follows. In Sec.~\ref{s2}, the 
experimental setup and procedure are discussed in detail. In 
Sec.~\ref{s3}, we use and compare different methods for 
identification of the ``solidlike'' particles that comprise the 
crystal domains and isolate the domain surface. In 
Sec.~\ref{s4}, the rectilinear propagation definition of the 
crystallization front velocity is proposed. The data on the 
front velocity measurement are discussed in Sec.~\ref{s5}. 
The results of this study are summarized in Sec.~\ref{s6}.

\section{\label{s2}EXPERIMENTAL SETUP AND PROCEDURE}

\begin{figure}
	\includegraphics[width=0.95\columnwidth]{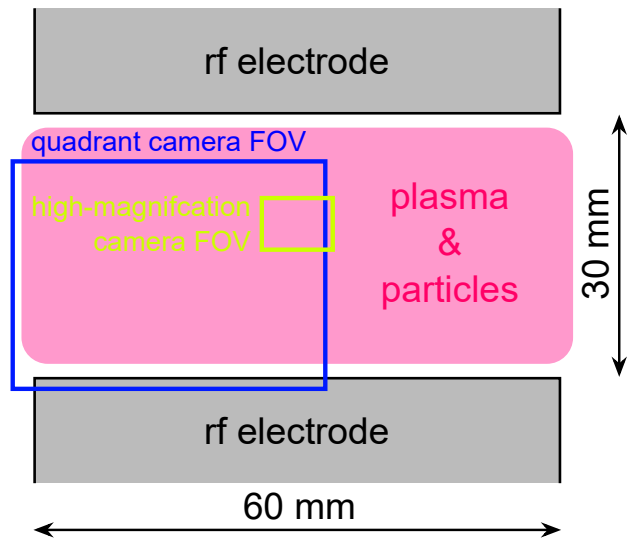}
	\caption{\label{f1}
		Sketch of the PK-3 Plus plasma chamber\cite{18}. Fields of view (FOV) of different videocameras are shown.}
\end{figure}

In the PK-3 Plus laboratory the plasma is excited by an rf-generator of 4W maximum at 13.56MHz with a symmetrical output to drive both electrodes in a push-pull mode. 
This symmetrical coupling provides a symmetric distribution of the plasma between the two
electrodes. A schematic of the experiment is shown in Fig.~\ref{f1}. 
Details on the setup can be found in \cite{18}. 
The electrodes are aluminum disks with a diameter of $6\,{\mbox{cm}}$ and are $3\,{\mbox{cm}}$
 apart.

The optical particle detection system consists of a laser 
diode with optics that shapes a ``laser sheet'' 
perpendicular to the electrode surface and the video 
cameras. The progressive scan CCD-cameras observe the 
scattered light at $90^\circ$ with different magnifications 
and fields of view. The quadrant camera shows the left part 
of the interelectrode system (about half of the full system) 
with a field of view~(FOV) of $35.7 \times 26.0\,{\mbox{mm}}^2 
$. The highest magnification camera can be moved along 
the central vertical axis and has a FOV of $8.1 
\times 5.9\,{\mbox{mm}}^2$ and is used for 
high-precision measurement of the microparticles position. 
The fields of view of the two cameras are shown in 
Fig.~\ref{f1}. 
The cameras follow the PAL standard, have a resolution of $768 \times 576\,{\mbox{pixels}}$, 
and provide the two composite $25\,{\mbox{Hz}}$ interlaced video channels each.
We have mixed these two channels to a $50\,{\mbox{Hz}}$ progressive scan video.
The cameras and lasers are mounted on a horizontal translation stage allowing a depth scan through and, therefore, a 3D view of the complex plasma \cite{101,Thomas2019,Schwabe2018}.

For manipulation and excitation of the complex plasma, a 
low-frequency function generator was utilized. The 
function generator provides amplitudes at low frequencies 
to the electrodes, overlaid on the rf signal. Frequencies 
between $1$ and $255\,{\mbox{Hz}}$
 with amplitudes up to $ \pm 55\,{\mbox{V}}$
 can be set with different waveforms and even different 
phases between the electrodes \cite{18}.

\begin{figure}
	\includegraphics[width=0.95\columnwidth]{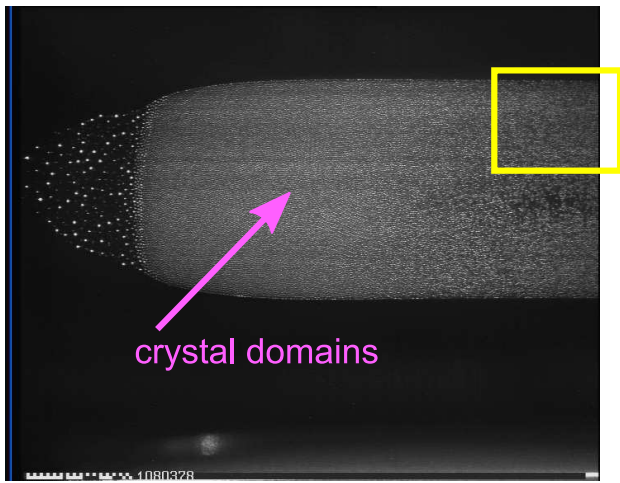}
	\caption{\label{f2}
		Side view of dust cloud obtained by the quadrant camera a short time after the function generator pulse. 
		The yellow rectangle shows the high-magnification camera FOV. 
		The purple arrow points to the region where the crystal domains have survived.}
\end{figure}

The experiment proceeded as follows. After formation of 
the plasma in argon, $1.55\,\mu {\mbox{m}}$
 diameter silica particles were injected into it, and the 
pressure was reduced to 10 or $15\,{\mbox{Pa}}$. 
The pressure was maintained constant in the course of the experiment. 
The measured effective rf-elecrode voltage was equal to 14.9~V and 14.4~V for 10 and 15 Pa,  respectively. 
Under these conditions, due to the intense flux of ions 
streaming from the discharge center, a space free from micro  
particles (void) is formed in the vicinity of the cloud center. 
In spite of the void, almost uniform microparticle number 
density $n_d = 7.4 \times 10^5 \,{\mbox{cm}}^{ - 3}$ in 
the high-magnification camera FOV is established. 
Then, the function generator was turned on at the frequency 
$255\,{\mbox{Hz}}$
(that is above the dusty plasma frequency which according to the estimation for the experimental conditions is less than 100 Hz)
and the voltage of $13\,{\mbox{V}}$ for $10\,{\mbox{s}}$.
The result of low frequency excitation is a drastic change of the plasma parameters.
The evidence of this is the void behavior which is nearly disappeared (like in \cite{18}) and the dust cloud turned into a highly homogeneous system in the high-magnification camera FOV.
Next, the function 
generator was turned off for $5\,{\mbox{s}}$
 and then it was turned on again at the voltage of 
$20\,{\mbox{V}}$. 
Note that this value of the voltage is less than the threshold voltage at which string formation takes place. 
In addition, the threshold voltage is greater for smaller particle sizes \cite{Ivlev2008}, i.e.\ the conditions of our experiments are far from those of the string formation.
As a result of such manipulation, the initial dust crystal was brought into a liquidlike state in FOV of the high-magnification camera. 
Thus, the function generator was used here for preparation of a 
``supercooled'' system, since the kinetic energy pumped 
into the microparticles dissipates in less than a second due 
to the Epstein neutral gas drag. Note that some minor 
crystal domains survive ``shaking'' by the function 
generator. However, these domains lie outside the view of 
the high-magnification camera, and one can see some 
domains by the ``quadrant'' camera (Fig.~\ref{f2}) that 
has a wider FOV. Then these peripheral crystal 
domains start to grow in all directions and, eventually, they 
appear in FOV of the high-magnification 
camera. The domain growth was observed with and without 
depth scans by the high-magnification camera. Nine 
forth and nine back depth scans were performed. The depth 
and the velocity ${u}$ of each scan were $4.8\,{\mbox{mm}}$
 and $0.6\,{\mbox{mm~s}}^{-1}$, respectively. Thus, one scan 
takes $8\,{\mbox{s}}$. The observation time between two 
successive scans was $3.7\,{\mbox{s}}$. Thus, the scans 
period was $23.4\,{\mbox{s}}$.
Note that we also performed the crystallization without the low frequency excitation.
We observed that the front propagation was too fast and the system in FOV was crystallized within a single scan period, which prevented the observation of the front propagation.

The video images recorded by the high-magnification 
camera allowed one to observe appearance and growth 
of the crystallization domains within FOV of 
this camera. The motion of the domain boundary 
illuminated by the laser sheet was observed in experiments 
without scans. With this purpose, the sequences of $100$
 video frames were superimposed in separate images with 
the interval of $10$
 frames, i.e., the resulting images are combinations of the 
initial frames from $1$ to $100$, from $10$
 to $110$, etc. In these images, the particles forming a 
crystal remain confined in their cells while the liquid 
particles move at about the interparticle distance. Thus, the 
particles belonging to the crystal form bright spots while 
the traces of particles from liquid are blurred over the 
frame, which makes the crystallization front visible 
(Fig.~\ref{f3}). Although this figure is a good illustration 
of the crystallization process, a two-dimensional (2D) video 
is insufficient for the determination of the 3D 
crystallization front propagation velocity, in particular, due 
to the fact that the velocity vector can lie outside the 
observation plane. For this purpose, we recovered the front 
surfaces at different instances from the depth scan data.

\section{\label{s3}IDENTIFICATION OF THE CRYSTALLIZATION FRONT}

\begin{figure}
\includegraphics[width=0.95\columnwidth]{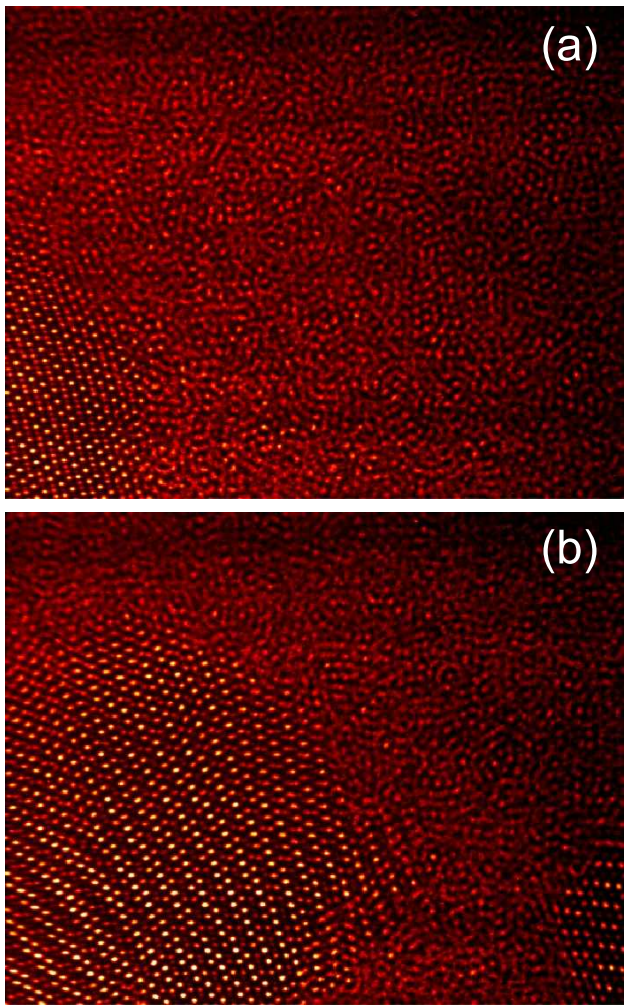}
	\caption{\label{f3}
		Demonstration of the liquidlike metastable dust cloud crystallization. 
		Each  image is a superposition of 100 consecutive frames (see text); phases (a) and (b) are separated by ca.\ $20\,{\mbox{s}}$.}
\end{figure}

The high-magnification camera FOV and 
scan depth restrict the volume accessible for investigation. It 
contains over $150000$
 particles. We employed the method of 3D particle 
coordinate determination \cite{69}. In what follows, we use 
the coordinate system with the $Z$-axis being the 
symmetry axis directed from the discharge center to an 
electrode. The $X$-axis lies in the plane of the laser sheet 
while the $Y$-axis is perpendicular to this plane. The 
direction of the latter coincides with the direction of odd 
scans. The coordinate system origin finds itself in the center 
of the discharge. One can conclude from the video 
illustrated by Fig.~\ref{f3} that the first domain emerges in 
the volume of observation at the seventh scan and then the 
volume occupied by the crystal phase increases until the 
$11$-th scan, where all domains cease to grow. Further, the 
whole volume is filled with adjoined crystal domains (but 
the near-sheath region, where crystallization is impossible). 
Hence, it is sufficient to treat the data from scan \#7 to 11.

The first task is the identification of ``solidlike'' and 
``liquidlike'' particles followed by isolation of individual 
domains. The widely used method, the correlation criterion 
is based on the introduction of the local order parameter 
\cite{93,96}
\begin{equation}
\bar q_{lm} (i) = \frac{1}{{N_i }}\sum\limits_{j = 
1}^{N_i } {Y_{lm} ({\bf{\hat r}}_{ij} )} , \label{e1}
\end{equation}
where $Y_{lm}$ are the spherical harmonics\cite{Klumov2014}, ${\bf{\hat 
r}}_{ij} = ({\bf{r}}_j - {\bf{r}}_i )/\left| {{\bf{r}}_j - 
{\bf{r}}_i } \right|$
 is the unit vector that specifies the direction from the 
particle $i$
 to its neighbor $j$, ${\bf{r}}_i$ and ${\bf{r}}_j$ are the 
radius-vectors of the respective particles, and the sum runs 
over all $N_i$ nearest neighbors of the particle $i$. Here 
and in what follows, we use the SANN algorithm \cite{97} 
for the definition of the nearest neighbors.

The local order parameter (\ref{e1}) makes it possible to 
define the correlation coefficient for a pair of particle $i$
 and $j$
\begin{equation}
\kappa _{ij} = \frac{{\sum\limits_{m = - 6}^6 {\bar 
q_{6m} (i)\,\bar q_{6m} (j)^* } }}{{\left[ {\sum\limits_{m 
= - 6}^6 {\left| {\bar q_{6m} (i)} \right|^2 } 
\sum\limits_{m = - 6}^6 {\left| {\bar q_{6m} (j)} \right|^2 
} } \right]^{1/2} }}, \label{e3}
\end{equation}
where asterisk denotes complex conjugation. According to 
\cite{96} we identify the particle $i$
 as a ``solidlike'' one if it has at least $11$
 nearest neighbors, for which $\left| {\kappa _{ij} } \right| > 
0.5$
 (in this case, we call the particles $i$
 and $j$
 highly correlated). Such correlation criterion proved to be a 
powerful tool in separation of individual particles into the 
group of ``solidlike'' and ``liquidlike'' ones. Thus, one can 
effectively define the interface between the phases. 
However, this criterion is incapable of identification of the 
interface between different crystal domains and hence, it 
cannot isolate them. Instead, the objective of this study is 
investigation of the interface including its parts contacting 
with other domains. Thus, it is very important for our 
problem to develop an alternative algorithm that is capable 
of identifying both particles and domains.

The idea of the algorithm we used is based on the fact that 
the orientation of the crystal axes is different for different 
domains. Thus, one can distinguish whether a particular 
particle belongs to a given domain if the directions to its 
nearest neighbors are close to the directions of the crystal 
axes. Bearing this in mind, for each particle, we try to find 
pairs of almost opposite nearest neighbors, which define the 
orientation of the unit cell. Then we do the same for one of 
the neighboring particles and check whether the next unit 
cell has the same orientation. Due to slight distortions, the 
directions of all pairs may not coincide, so we restrict the 
number of coincidences to five. If this is the case, we 
identify both particles as solidlike ones pertaining to the 
same domain.

We will term this the ``axis'' algorithm. It is realized in the 
following way. We choose the $k$-th particle with the 
radius-vector ${\bf{R}}_k$ at random and find all $N_k$ 
nearest neighbors. At this step, we consider the $k$-th 
particle as a ``central'' one belonging to the treated domain. 
Let ${\bf{R}}_l = {\bf{r}}_i^{(k)}$ be the radius-vector 
of one of its nearest neighbors, where $l$
 is the numeration of this particle in the whole system. 
 As a local characteristic of the crystalline lattice, we 
 introduce the matrix $\rho _{ij}^{(k)} = \left| 
 {{\bf{r}}_i^{(k)} + {\bf{r}}_j^{(k)} - 2{\bf{R}}_k } 
 \right|$. We will term counterparts the two nearest 
 neighbors defined by the indexes $i_0$ and $j_0$ if for 
 fixed $i = i_0 $, $\rho _{i_0 j}^{(k)}$ is minimized at $j = 
 j_0$ as $j$ runs from $1$ to $N_k$ and, reciprocally,
 for fixed $j = j_0 $, $\rho _{ij_0 }^{(k)}$ is minimized
 at $i = i_0$ as $i$ runs from $1$ to $N_k $.
 If an additional condition $\rho _{ij}^{(k)} < \rho 
_c $, where $\rho _c$ is some small distance parameter, is 
satisfied then we define the unit vector 
${\bf{d}}_{ij}^{(k)} = \left( {{\bf{r}}_i^{(k)} - 
{\bf{r}}_j^{(k)} } \right) \left| {{\bf{r}}_i^{(k)} - 
{\bf{r}}_j^{(k)} } \right|^{-1}$. For the counterparts in an ideal 
lattice, $\rho _{ij}^{(k)} = 0$, and ${\bf{d}}_{ij}^{(k)}$ 
defines the direction of a crystal axis. Thus, we form a set 
of the vectors ${\bf{d}}_{ij}^{(k)}$ corresponding to all 
found pairs of the counterparts. Then, this procedure is 
repeated for the nearest neighbor with the radius-vector 
${\bf{R}}_l $, and a set of the vectors 
${\bf{d}}_{ij}^{(l)}$ is generated. If at least $\eta = 5$
 pairs of the vectors ${\bf{d}}_{ij}^{(k)}$ and 
${\bf{d}}_{ij}^{(l)}$ satisfy the condition
\begin{equation}
\left| {{\bf{d}}_{ij}^{(k)} \cdot {\bf{d}}_{ij}^{(l)} } 
\right| > 1 - \varepsilon , \label{e4}
\end{equation}
where $\varepsilon > 0$
 is a parameter, then by definition, this neighbor belongs to 
the same domain. Otherwise, this neighbor is ignored in 
what follows. In so doing, we compare the local directions 
of the crystal lattice vectors (or antiparallel ones), which 
were determined most reliably.

For an ideal lattice, the corresponding pairs would satisfy 
the relation $\left| {{\bf{d}}_{ij}^{(k)} \cdot 
{\bf{d}}_{ij}^{(l)} } \right| = 1$. For our system, the 
optimum parameters that provide the sharpest domain 
identification proved to be $\rho _c = 50\,\mu 
{\mbox{m}}$
 and $\varepsilon = 9.73 \times 10^{ - 3} $. The parameter 
$\rho _c$ is still much less than the doubled mean 
interparticle distance $\left| {{\bf{r}}_i^{(k)} - 
{\bf{r}}_j^{(k)} } \right| \sim 2n_d^{ - 1/3} = 221\,\mu 
{\mbox{m}}$, and the choice of $\varepsilon$ ensures that 
the least angle between the vectors ${\bf{d}}_{ij}^{(k)}$ 
and ${\bf{d}}_{ij}^{(l)}$ or between 
${\bf{d}}_{ij}^{(k)}$ and $ - {\bf{d}}_{ij}^{(l)}$ is less 
than $8^\circ $. Obviously, these parameters can be 
sensitive to the lattice parameters.

At the next step, we choose the next nearest neighbor until 
all of them are checked. Likewise, the particles that satisfy 
condition (\ref{e4}) are marked as the ones belonging to the 
treated domain. Otherwise, they are marked as non-domain 
particles. However, they are still taken into account as the 
nearest neighbors for a ``central'' particle other than the 
$k$-th particle and, in principle, they can be added to the 
domain provided that the condition (\ref{e4}) is satisfied 
for them. Then, we consider one of the neighbor particles as 
a ``central'' one and repeat the entire procedure, etc., until 
all the particles that form the domain are detected. At the 
next stage, we select randomly the next initial particle not 
belonging to any domain and find a new domain until all of 
them are detected.

\begin{figure}
	\includegraphics[width=0.95\columnwidth]{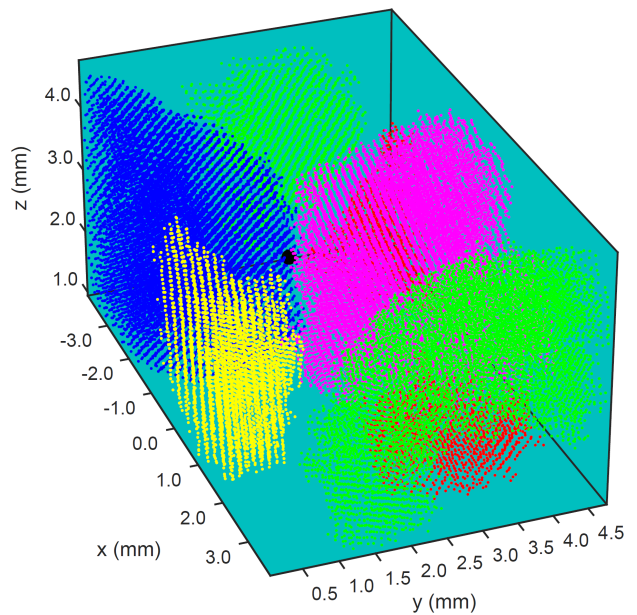}
	\caption{\label{f4}
		Domains of the true ``crystallike'' particles for scan \#11 resolved on the basis of the ``axis'' algorithm. 
		Color indicates particular orientation of the crystal axes. 
		The ``liquidlike'' particles and the crystal domains containing less than $1000$ particles are not shown. 
		Black circle at the interface of the blue-colored domain indicates the intersection with the line along which the crystallization front propagation is treated (the line of rectilinear propagation).}
\end{figure}

A comparison between this ``axis'' algorithm and the 
correlation criterion shows that both methods mark almost 
the same particles as ``solidlike'' ones if $\eta$ is properly 
adjusted ($\eta < 5$). However in this case, the ``axis'' 
algorithm is incapable of separation of different domains as 
well as the correlation criterion. The above-selected 
parameter $\eta = 5$
 is the minimum one that solves the problem of separation. 
Albeit the number of ``solidlike'' particles in this case is 
more than 25\% lower than that resolved by the correlation 
criterion (for scan \#7), the mismatch takes place mainly 
near the domain surface and the planes that confine the 
camera FOV and scan depth (see Fig.~\ref{f4}). 
In addition, all particles defined as 
``solidlike'' by the ``axis'' algorithm are either labeled 
likewise by the correlation criterion or they are highly 
correlated. Note that the shape of the solid--liquid interface 
that depends weakly on the particle number proves to be 
little different for both methods. Figure~\ref{f4} shows 
individual crystal domains for scan \#11 resolved on the 
basis of the ``axis'' algorithm, for which the number of the 
``solidlike'' particles is maximum. The differently oriented 
crystal lattices are clearly seen in this figure.

\begin{figure}
	\includegraphics[width=0.95\columnwidth]{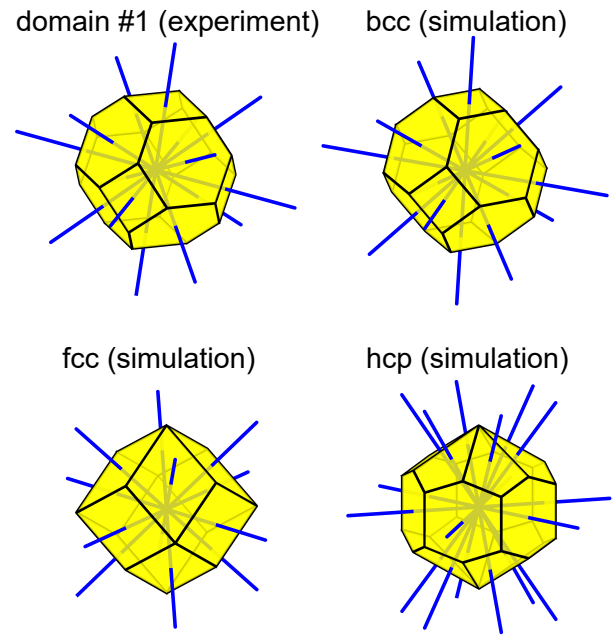} 
	\caption{\label{f5}
	Voronoi cells calculated from the virtual lattices for the experimental domain (the blue-colored domain in Fig.~\ref{f4}) and the simulated bcc, fcc, and hcp crystalline lattices.}
\end{figure}

It is of interest to determine the type of the crystalline structure for the treated domains. 
For this purpose, we can apply the axis method described above. 
Due to the approximate periodicity of a real lattice, the points in space 
defined by the vectors ${\bf{r}}_l^{(k)} - {\bf{R}}_k $
that define the coordinates of the $l$th particle relative to the coordinate system with the origin 
at the location of the $k$th particle ($l$
and $k$
enumerate the particles in a whole domain) form clusters. 
We determine the centers-of-mass ${\bf{\bar R}}_m $, 
where $m$
enumerates clusters, for such clusters using the $K$-means 
algorithm \cite{Jain2010}. The points ${\bf{0}}$
and ${\bf{\bar R}}_m$ form a virtual ``averaged'' lattice, 
which, by and large, does not coincide with a real one due 
to the lattice defects, thermal motion of the particles, and 
the peculiarities of a crystalline symmetry (e.g., for the hcp 
type lattice, such virtual lattice is quite different from hcp). 
Then we perform the Voronoi tessellation of this virtual 
lattice and isolate a single Voronoi cell around the point 
${\bf{0}}$. Thus, we obtain a polyhedron that can be 
indexed by the number and type of its faces 
(Fig.~\ref{f5}). 

We applied this algorithm, which will be discussed in detail 
elsewhere, to the simulated ``noisy'' bcc, fcc, and hcp 
lattices deformed by random shifts of the particles from 
their equilibrium positions in a crystalline lattice. We found 
that corresponding polyhedrons are fully independent of the 
shift amplitude provided that it is not too large and that they 
can be indexed as follows. The bcc polyhedron has six 
quadrangular and eight hexagonal faces; the fcc 
polyhedron, 12 quadrangular faces; and the hcp polyhedron, 
12 quadrangular and six hexagonal faces. 
Comparison between the indexes for the blue-colored 
domain in Fig.~\ref{f4} (domain \#1) and the simulated bcc 
reveals an exact match and a total mismatch with the 
simulated fcc and hcp, whence it follows that the lattice 
type is bcc. The magenta-colored domain in Fig.~\ref{f4} 
(domain \#2), to which the above-discussed procedure was 
applied, appeared to be of the same bcc type. Note that no 
evidence of the crystalline anisotropy follows from 
Fig.~\ref{f5}: the polyhedron seems almost symmetric. 
The absence of anisotropy under compatible experimental 
conditions was noted in \cite{94}. 
Note that all analyzed experimental domains (consisting of more than 100 particles) have the bcc structure. 
This is true for the experiment with the gas pressure of 15 Pa (domains \#3 and 4, see Sec.~\ref{s4}). In particular, the bcc structure has been observed in the dusty plasma studied at the ground-based facility \cite{Dietz2018} analogous to the PK-3 Plus facility.

For the further analysis, we select the domain \#1 as it is the 
largest one. At this stage, our objective is the 
definition of the crystallization front. If we neglect the 
fluctuations of the domain boundary on the interparticle 
distance scale and assume the rectilinear propagation of a 
front small area (see Sec.~\ref{s4}) then some arbitrariness 
is involved in the front definition. In this case, the distance 
between this front and the equimolar surface has no 
physical meaning, similar to the definition of the surface of 
tension for a plane interface. For example, it is not 
important whether highly correlated particles that have less 
than $11$
 highly correlated neighbors (mostly, the boundary 
particles) are included in the front definition. One only 
needs to use the same front definition for all treated scans. 
Hence, it is sufficient to isolate a monolayer of the particles 
at the surface of a domain that comprises only true 
``solidlike'' particles. This monolayer consists of pivot 
particles for the crystallization front. Isolation of the surface 
monolayer is performed using the algorithm \cite{98} that 
proved good at the Lennard-Jones cluster surface 
determination. According to it, the particle 1 with the 
radius-vector ${\bf{r}}_1$ that belongs to the domain will 
be called internal if there exists at least a single particle 2 
with the radius-vector ${\bf{r}}_2$ belonging to the same 
domain that has more than four nearest neighbors such that 
the conditions
\begin{equation}
{\bf{r}}_1 \cdot {\bf{r}}_2 > r_1^2 ,\quad r_2^2 - 
\frac{{({\bf{r}}_1 \cdot {\bf{r}}_2 )^2 }}{{r_1^2 }} < 
L^2 , \label{e5}
\end{equation}
are satisfied. Here, $L$
 is the length parameter on the order of the interparticle 
distance. The particles that are not internal and have more 
than four nearest neighbors are the surface (or pivot) 
particles.

\begin{figure}
	\includegraphics[width=0.95\columnwidth]{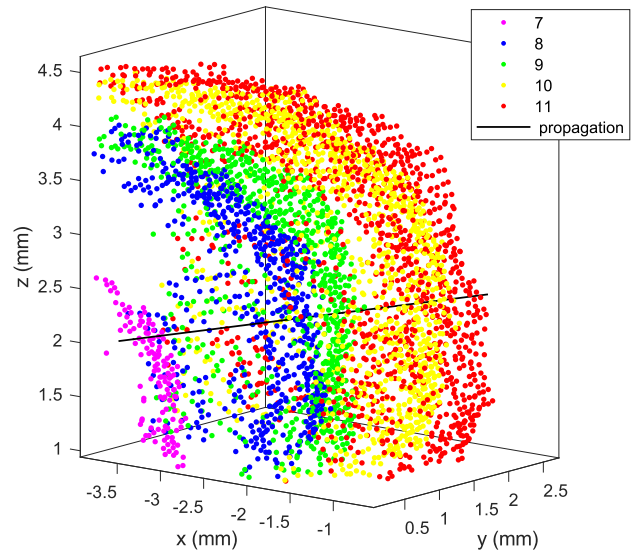}
	\caption{\label{f6}
		Development of the crystallization front in time for the blue-colored domain in Fig.~\ref{f4}. 
		Shown are the domain surfaces determined using (\ref{e5}) corresponding to scans \#7--11. 
		The scan numbers are color-coded (see legend). 
		Solid line indicates the front propagation direction.}
\end{figure}

For this definition, we shift the origin of the coordinate 
system to the point ($x = - 4\,{\mbox{mm}}$, $y = 0$, and 
$z = 0.9\,{\mbox{mm}}$). The l.h.s.\ of the second 
condition (\ref{e5}) is the squared distance between particle 
2 and the axis passing through the coordinate system origin 
and the particle 1. We set $L^2 = 0.45r_{\min }^2 $, where 
$r_{\min }$ is the distance between the particle 1 and its 
closest neighbor. The results of the definition of 
crystallization front for scans \#7--11 are shown in 
Fig.~\ref{f6}. Note that the selected algorithm effectively 
eliminates the particles adjacent to the boundaries of the camera FOV, which 
do not form any physical surface. It is seen that beginning 
with scan \#7, the surface of selected domain (blue-colored 
in Fig.~\ref{f4}) becomes visible and that this domain 
grows until it is confined by the boundaries of neighboring 
domains and by the near-electrode sheath. A little 
difference between the fronts for scans \#10 and 11 is 
indicative of the fact that the domain growth stops at this 
time. Obviously, the whole crystallization front has no 
definite shape. Since the front velocity is correctly 
determined only for a plane front propagating along a 
certain direction, we have to search for such a plane area of 
the front and for the corresponding direction of 
propagation.

\section{\label{s4}MEASUREMENT OF THE FRONT VELOCITY}

Measurement of the 3D crystallization front propagation 
velocity encounters two principal problems. First, the time 
between successive scans is long enough so that the path of 
a small front area can be curvilinear and therefore, it cannot 
be measured. This makes even a physical definition of the 
front velocity problematic. Second, the front velocity $v$
 cannot be fully neglected as compared with the scan 
velocity $u$. Hence, the visible domain surface determined 
in the depth scan is different from the real one. Both 
difficulties are obviated if we assume that there exists a 
small area of the domain surface that can be approximated 
by a plane and that propagates along a certain line with the 
normal to this area ${\bf{n}} = \left\{ {n_x ;\;n_y ;\;n_z } 
\right\}$
 parallel to this line (${\bf{n}}$
 is directed like the propagation velocity). We will term 
such small area the reference area and define the front 
propagation velocity $v$
 as the velocity of the reference area motion along 
${\bf{n}}$.

Consider a pair of even or odd scans, for which the visible 
reference areas are parallel. The normal to the visible 
reference area ${\bf{n'}}$
 can be related to ${\bf{n}}$. Consider a pair of the odd 
scans, 1 and 2, when the dust cloud is scanned in the 
direction of the $Y$-axis. Then the plane of the laser sheet 
during scan 1 is defined by the equation $y = u(t - \tau _1 
)$, where $t$
 is the time and $\tau _1$ designates the start time of scan 
1. The equation for the ``true'' reference area is ${\bf{n}} 
\cdot {\bf{r}} = vt$, where ${\bf{r}}$
 designates space coordinates. The visible reference area is 
a plane, in which the intersection between the laser sheet 
plane and the ``true'' reference area moves. We exclude the 
time from the equations defining the laser sheet plane and 
the reference area to obtain
\begin{equation}
{\bf{n'}} \cdot {\bf{r}} = r_1 = \frac{{v\tau _1 }}{\eta 
},\quad {\bf{n'}} = \left\{ {\frac{{n_x }}{\eta 
};\;\frac{{n_y - \beta }}{\eta };\;\frac{{n_z }}{\eta }} 
\right\}, \label{e6}
\end{equation}
where $r_1$ is the distance between the visible reference 
area and the origin of the coordinate system,
\begin{equation}
\eta = \sqrt {n_x^2 + \left( {n_y - \beta } \right)^2 + n_z^2 
} \simeq 1 - \beta n_y \simeq 1 - \beta n'_y \label{e7}
\end{equation}
and $\beta = v/u$, which is assumed to be small, $\beta \ll 
1$.

We approximate the visible reference area by a secant plane 
that is close to the tangential plane perpendicular to the 
normal ${\bf{n'}}$. We find a small group of $m$
 particles in the vicinity of the point of intersection between 
the front propagation line and the domain surface with the 
radius-vectors ${\bf{r}}_i$ that maximize the dot product 
${\bf{r}}_i \cdot {\bf{n'}}$. The radius-vector of the 
center of mass of the selected group is
\begin{equation}
{\bf{R}}_c^{(1)} = \frac{1}{m}\sum\limits_{i = 1}^m 
{{\bf{r}}_i } \label{e79}
\end{equation}
and its projection on the direction of ${\bf{n'}}$
 is
\begin{equation}
r_1 = {\bf{n'}} \cdot {\bf{R}}_c^{(1)} = \frac{1}{m}
\sum\limits_{i = 1}^m {{\bf{r}}_i \cdot {\bf{n'}}} . 
\label{e8}
\end{equation}
It follows from Eqs.~(\ref{e6}) and (\ref{e7}) that
\begin{equation}
\beta u\tau _1 = \left( {\beta n'_y - 1} \right)r_1 . \label{e9}
\end{equation}
For the next odd scan, $\beta u\tau _2 = \left( {\beta n'_y - 
1} \right)r_2 $, where $\tau _2$ is the start time of scan 2 
and $r_2$ is defined by the r.h.s.\ of Eq.~(\ref{e8}) for scan 
2. We subtract (\ref{e9}) from this equation to derive the 
front velocity between the successive odd scans,
\begin{equation}
v = v_{\mathrm{odd}} = \frac{{v_0 }}{{1 + n'_y 
\frac{{v_0 }}{u}}},\quad v_0 = \frac{{r_2 - r_1 }}{{\tau 
_2 - \tau _1 }}. \label{e10}
\end{equation}
As is seen from (\ref{e10}), the front velocity is invariant to 
the coordinate system and depends solely on the difference 
of the start times of like scans. The equation for the front 
velocity between successive even scans differs from 
(\ref{e10}) by the opposite sign of $u$,
\begin{equation}
v = v_{\mathrm{ev}} = \frac{{v_0 }}{{1 - n'_y \frac{{v_0 
}}{u}}}. \label{e11}
\end{equation}
The last term in the denominator of (\ref{e10}) and 
(\ref{e11}) is the correction for a finite scan velocity.

Since the rectilinear propagation of the reference area 
perpendicular to the line of propagation can be 
characterized by the coincidence of directions of the vector 
${\bf{R}}_c^{(2)} - {\bf{R}}_c^{(1)}$ that defines the 
propagation line and the ``true'' normal
\begin{equation}
{\bf{n}} = \{ (1 - \beta n'_y)n'_x;\;(1 - \beta n'_y)n'_y + \beta;\; (1 - \beta n'_y)n'_z\}, \label{e12}
\end{equation}
we have to minimize the difference between 
${\bf{R}}_c^{(2)} - {\bf{R}}_c^{(1)}$ and its projection 
on ${\bf{n}}$,
\begin{equation}
\rho _c = \left| {{\bf{n}} \times \left( {{\bf{R}}_c^{(2)} - 
{\bf{R}}_c^{(1)} } \right)} \right|, \label{e129}
\end{equation}
by variation of ${\bf{n'}}$. Here, we ignore unphysical 
boundary value maxima corresponding to the propagation 
directions that are almost parallel to the boundaries of the camera FOV. At 
the same time, we check that the reference area is not a part 
of the interface between two domains.

Thus, the front propagation velocity is determined as 
follows. For every treated domain, we resolve its visible 
surface at each scan, then we select relevant pairs of the 
consecutive odd-odd or even-even scans. For a pair of 
scans, we adjust ${\bf{n'}}$
 to minimize $\rho _c$ (\ref{e129}) using formulas 
(\ref{e79}) and (\ref{e12}). In this way, we determine both 
the propagation direction and velocity (Eqs.~(\ref{e8}), 
(\ref{e10}), and (\ref{e11})).

\begin{figure}
	\includegraphics[width=0.95\columnwidth]{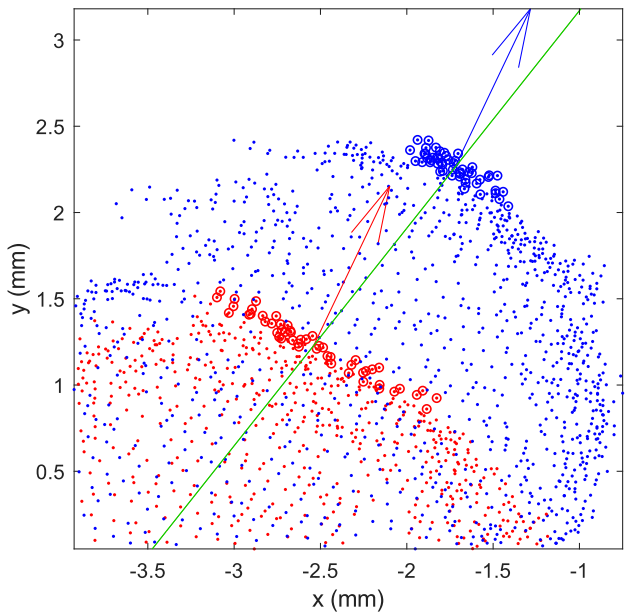}
	\caption{\label{f7}
		Illustration of the rectilinear propagation front velocity measurement for scans \#8 and 10 (blue-colored domain in Fig.~\ref{f4}). 
		Red dots indicate the domain surface for scan \#8 and blue dots, for scan \#10; circled dots show the particles that comprise the reference areas. 
		Green line indicates the propagation direction parallel to the ``true'' normal ${\bf{n}}$; red and blue arrows show the vector ${\bf{n'}}$	starting at the centers of mass of the reference areas for corresponding scans.}
\end{figure}

Figure~\ref{f7} illustrates the results of minimization 
procedure for the pair of scans \#8 and 10. It is seen that 
the difference between ${\bf{n'}}$
 and ${\bf{n}}$
 (propagation direction) is noticeable, so that the correction 
for the finite scan velocity is appreciable. We have found 
that the optimum number of particles comprising the 
reference area is $m = 50$
 and the resulting $\rho _c = 23\,\mu {\mbox{m}}$
 while the interparticle distance is $n_d^{ - 1/3} = 111\,\mu 
{\mbox{m}}$. The condition $\rho _c n_d^{1/3} \ll 1$
 satisfied for our system is indicative of the fact that there 
exists at least one rectilinear propagation direction and 
corresponding reference area.

We applied the above-discussed procedure for the two 
largest domains, namely, the blue-colored and 
magenta-colored ones in Fig.~\ref{f4}. Other domains have 
too small surface at the scans \#7--10 to be processed. For 
the blue-colored domain, the pair of scans \#8 and 10 is a 
single one that satisfies our conditions. Indeed, for the odd 
pair \#7 and 9, the propagation line passes outside the 
domain surface at scan \#7, which is partly invisible 
(Fig.~\ref{f6}). For the same reason, we failed to find a 
non-boundary minimum of $\rho _c$ for this scan pair. In 
the last pair of scans (\#9 and 11), the reference area for 
scan \#11 is a boundary between two neighboring domains 
(Fig.~\ref{f4}). For this reason, this pair is unusable. 
However, given the propagation line determined for scans 
\#8 and 10 (black line in Fig.~\ref{f6}), one can calculate 
the distance between its intersections with the domain 
surfaces (the front displacement) for all possible scan pairs 
as $\Delta S = {\bf{n}} \cdot \left( {{\bf{r}}_2 - 
{\bf{r}}_1 } \right)$, where ${\bf{r}}_2$ and 
${\bf{r}}_1$ are the radius-vectors of the particles 
pertaining to the surface of corresponding domain that are 
situated at the closest distance from the propagation line. 
The coordinates $z$
 of such particles are unambiguously related to the instants 
$\tau _2$ and $\tau _1 $, at which they are illuminated by 
the laser sheet. Note that there is no real point of 
intersection between the propagation line and the scan \#7 
surface. The corresponding time difference $\Delta \tau = 
\tau _2 - \tau _1$ enables determination of the front 
velocity by such point measurement, $v = \Delta S/\Delta 
\tau $. For the magenta-colored domain in Fig.~\ref{f4}, 
the point measurement is redundant because all three pairs 
of scans, \#7 and 9, \#8 and 10, and \#9 and 11 can be 
processed. 

We performed similar experiment for the argon pressure of 15 Pa. In this case, we also found the two largest domains with sufficient area of the solid--liquid interface (domains \#3 and 4). The rectilinear propagation way of the velocity measurement is applicable to the pairs of scans \#9--11, \#10--12, \#11--13, and \#12--14 (domain \#3) and \#11--13 and \#12--14 (domain \#4).

\section{\label{s5}RESULTS AND DISCUSSION}

\begin{figure}
	\includegraphics[width=0.95\columnwidth]{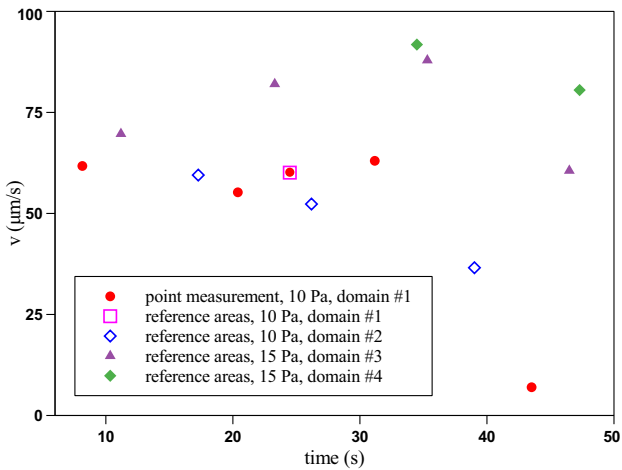}
	\caption{\label{f8}
		Crystallization front propagation velocity determined from the rectilinear propagation velocity definition for the domain \#1 (open square), \#2 (open diamonds), \#3 (triangles), and \#4 (solid diamonds), and by the dot measurement for the domain \#1 (circles). 
		The domains \#1 and 2 correspond to the experiment with the argon pressure of 10 Pa, the domains \#3 and 4, to 15\,Pa. The time is counted from the start of a scan for which the domains \#1 and 3 are first visible in the camera FOV for the gas pressure of 10 and 15\,Pa, respectively.}
\end{figure}

The minimization procedure discussed in Sec.~\ref{s4} 
yields the estimation for the front velocity $v \simeq 
59\,\mu {\mbox{m~s}}^{-1}$. The results of measurements 
performed for four domains are summarized in 
Fig.~\ref{f8}. This figure is indicative of the fact that the 
velocity of crystallization front is almost independent both 
of the propagation time and of the spatial position in the 
dust cloud. Note that a sharp drop of the velocity after 
$40\,{\mbox{s}}$
 is accounted for by the contact between the boundaries of 
neighboring domains. This contact can increase the tension 
inside the domains and thus shift them toward the 
crystallization--melting binodal. 
A good agreement between the velocities 
determined by the minimization procedure for all 
domains and the point measurement is worth mentioning 
(Fig.~\ref{f8}). 
The results for the argon pressure of 15\,Pa are not much different from that for 10\,Pa. It is seen in Fig.~\ref{f8} that the front velocity is somewhat higher than that for 10\,Pa and reveals a low maximum followed by the velocity fall. 
However for both pressures, most of the data are grouped in the interval from $60$ to $80\,\mu {\mbox{m~s}}^{-1}$. Note that for 15\,Pa, as domains become visible in the camera FOV, they contact each other so that the surface area of the domain--domain interface is substantial. 
Extensive contacts between the domains can modify the front velocity but the surface area of such contacts can hardly be measured in experiments. 
In contrast, the measurements using the domains\,\#1 and 2 correspond to the growth of free domains, which increases the reliability of these data.
Since for different domains, the front propagation directions are different, one can conclude that the front propagation velocity is independent of the direction, in which it is measured.

Apparently, the Coulomb coupling parameter is the most important one that can affect the front velocity. For the treated dust cloud, we estimate it using the results of our 
previous study \cite{94},
\begin{equation}
\Gamma = \frac{{Z^2 e^2 }}{{r_d k_B T_d }} = 3\left( 
{\frac{{r_d }}{{\delta r}}} \right)^2 , \label{e16}
\end{equation}
where $Z$ is the dust particle charge in units of the electron charge, $e$  is the elementary electric charge, $k_B$ is the Boltzmann constant, $T_d = 2K/3k_B$, and $K$ are the kinetic temperature and energy of the particles, respectively,  
$r_d = (3/4\pi n_d )^{1/3}$ is the Wigner--Seitz radius for 
a dust particle, and $\delta r$
 is the 3D standard deviation of a particle from its 
equilibrium position in this cell. The Coulomb coupling 
parameter measured for the runs without the depth scans 
proves to be almost constant in the region of front 
propagation  and amounts to 54 for 10\,Pa and to 49 for 15\,Pa. 
Such a low value seems to be meaningful because for $2.55\,\mu 
{\mbox{m}}$
 diameter particles, $\Gamma \sim 150$
 and it drops with the decrease of the particle diameter 
\cite{94}. 
Thus, we can conclude that we have measured the 3D velocity of 
the crystallization front propagating in complex plasma. 
This result is compatible with that of study \cite{89}, where 
the estimated vertical 2D velocity of the crystallization 
front was about one interparticle distance per second.

\section{\label{s6} CONCLUSION}

In this study, we have investigated the kinetics of the 
first-order phase transition in complex plasma, namely, the 
propagation of the crystallization front through the cloud of 
charged dust particles under microgravity conditions. Based 
on PK-3 Plus laboratory onboard the ISS, we used the 
function generator to prepare a ``supercooled'' state of the 
liquidlike dust cloud in plasma of a low-pressure argon 
discharge. We observed growth of the crystal domains in 
the dust cloud. The data obtained from the depth scans of 
this particle subsystem enabled us to obtain the 3D particle 
coordinates. Then we tried two approaches that can divide 
all particles into ``liquidlike'' and ``solidlike'' ones. 
The correlation criterion related to the rotational invariants 
can effectively distinguish between the particle types. 
However, it cannot separate different crystal domains and 
their surfaces. For this reason, we put forward the 
determination of the principal axes directions for each 
crystal that comprises a domain. This enables us not only to 
identify the ``solidlike'' particles but to separate effectively 
different domains. Comparison between these approaches 
shows that they identify almost the same particles as the 
``solidlike'' ones. The method of the domain surface 
determination we used allows one to isolate a set of the 
pivot particles that represent directly the propagating front.

A key point in the front velocity measurement is 
determination of the propagation line direction. This 
implies that there exists a small area of the domain surface 
propagating along this line perpendicularly to it that we 
term the reference area. We find such line by minimization 
of the projection of the vector connecting the reference area 
centers of the corresponding unidirectional scans on the 
normal to these areas. Here, we take into account rotation 
of the visible normals due to the finite scan velocity. In the 
course of this procedure, the crystallization front velocity is 
calculated. This is a true 3D front velocity that is an 
important property of the plasma crystallization kinetics. 
Alternatively, one can calculate the velocity dividing the 
distances between the points of intersection between the 
propagation line and the domain surfaces by the 
corresponding time intervals.

Eventually, we have found that the front velocity is almost 
constant for our system and it amounts to ca.\ 60--80\,$\mu 
{\mbox{m~s}}^{-1}$, which is less than the interparticle 
distance per second. A good agreement between the point 
measurement and the rectilinear propagation velocity 
definition is noteworthy. We have shown that the front 
propagates in a uniform dust cloud and it is decelerated 
when the surfaces of neighboring domains are brought in 
contact.

2D  measurements  show  solely  the line of the boundary of the domain
cross-section by the laser sheet plane. As the domain grows, different
points  of its surface intersect with the laser sheet. We suppose that
it  is  impossible to find a physical definition of the front velocity
based on the 2D measurements. Instead, the proposed 3D definition of this
quantity along the direction of the rectilinear propagation
is  definite  and it is a relevant characteristic of the corresponding
physical process.

The problems to be addressed in future are development 
of a theory of the plasma crystal growth and treatment of 
very small crystal clusters in the liquid (crystallites), which 
were neglected in this study. Investigations in these trends 
can contribute to development of the theory of strongly 
coupled plasma crystallization.

\begin{acknowledgments}
The PK-3 Plus laboratory project was supported by ROSCOSMOS. 
This project was also funded by Deutsches Zentrum f\"{u}r Luft- und Raumfahrt e.V. with funds from the Federal Ministry for Economy and Technology according to a resolution of the Deutscher Bundestag under grant numbers 50WM0203 and 50WM1203.
\end{acknowledgments}

\section*{References}
\providecommand{\noopsort}[1]{}\providecommand{\singleletter}[1]{#1}%

\end{document}